\documentclass[conference]{IEEEtran}

\usepackage[noadjust]{cite} % noadjust prevents cite from adjusting space before citation
\usepackage{graphicx}
\usepackage{amsmath}

\usepackage{tabularx} % For better table formatting, especially with text wrapping

\usepackage[hyphens]{url} % Allow line breaks at hyphens in URLs
\usepackage{hyperref} % Optional: for clickable links and PDF metadata
\hypersetup{
    colorlinks=true,       % false: boxed links; true: colored links
    linkcolor=blue,        % color of internal links (sections, pages, etc.)
    citecolor=green,       % color of links to bibliography
    filecolor=magenta,     % color of file links
    urlcolor=cyan,         % color of external links
    pdftitle={Securing Agentic AI: A Comprehensive Threat Model and Mitigation Framework for Generative AI Agents}, % PDF title
    pdfpagemode=FullScreen,% PDF view mode
    pdfauthor={Vineeth Sai Narajala, Om Narayan} % Add PDF author info
}

\IEEEoverridecommandlockouts
\begin{document}
\title{Securing Agentic AI: A Comprehensive Threat Model and Mitigation Framework for Generative AI Agents}

% Use the standard IEEE format
\author{
\IEEEauthorblockN{Vineeth Sai Narajala\textsuperscript{1} \thanks{\textsuperscript{1}This work is not related to the author’s position at Amazon Web Services.}}
\IEEEauthorblockA{\textit{Proactive Security} \\
\textit{Amazon Web Services}}
\and

\IEEEauthorblockN{Om Narayan\textsuperscript{2} \thanks{\textsuperscript{2}This work is not related to the author’s position at Amazon Web Services.}} % Shares the same footnote as Idan
\IEEEauthorblockA{\textit{Proactive Security} \\
\textit{Amazon Web Services} }
}

\maketitle

\begin{abstract}
As generative AI (GenAI) agents become more common in enterprise settings, they introduce security challenges that differ significantly from those posed by traditional systems. These agents aren’t just LLMs—they reason, remember, and act, often with minimal human oversight. This paper introduces a comprehensive threat model tailored specifically for GenAI agents, focusing on how their autonomy, persistent memory access, complex reasoning, and tool integration create novel risks. This research work identifies 9 primary threats and organizes them across five key domains: cognitive architecture vulnerabilities, temporal persistence threats, operational execution vulnerabilities, trust boundary violations, and governance circumvention. These threats aren’t just theoretical—they bring practical challenges such as delayed exploitability, cross-system propagation, cross system lateral movement, and subtle goal misalignments that are hard to detect with existing frameworks and standard approaches. To help address this, the research work present two complementary frameworks: ATFAA (Advanced Threat Framework for Autonomous AI Agents), which organizes agent-specific risks, and SHIELD, a framework proposing practical mitigation strategies designed to reduce enterprise exposure. While this work builds on existing work in LLM and AI security, the focus is squarely on what makes agents different—and why those differences matter. Ultimately, this research argues that GenAI agents require a new lens for security. If we fail to adapt our threat models and defenses to account for their unique architecture and behavior, we risk turning a powerful new tool into a serious enterprise liability.
\end{abstract}


% \fi
\IEEEpeerreviewmaketitle

\section{Introduction}
Generative AI (GenAI) agents are emerging as a new category of enterprise technology. Unlike conventional systems, they combine large language models (LLMs) with planning capabilities, persistent memory access, and third party/internal tool integration \cite{ref1}. These agents aren’t limited to generating responses—they actively interact with systems, make decisions, and act across enterprise environments, often with minimal human oversight \cite{ref2}.

This growing autonomy is what sets them apart—and what makes them particularly challenging from a security standpoint. GenAI agents can traverse organizational boundaries, make mutating API calls, and manipulate enterprise data, sometimes without direct user input \cite{ref3}. They’re dynamic, adaptive, and deeply embedded into operational workflows.

Traditional security measures may not fully address the risks these agents pose. Agentic architecture—a blend of reasoning components, memory systems, language interfaces, and external tools—introduces a much broader and more complex attack surface than most existing frameworks were designed to handle \cite{ref5}. While valuable, frameworks such as the OWASP Top 10 for LLMs \cite{ref9}, NIST AI Risk Management Framework \cite{ref2}, MITRE ATLAS \cite{ref6}, and CSA MAESTRO \cite{ref10} tend to treat LLMs as isolated components or provide high-level risk guidance. They often don’t account for the emergent security properties that arise when autonomy, long-term memory access, and dynamic tool usage are combined.

This paper aims to bridge that gap. It introduces the Advanced Threat Framework for Autonomous AI Agents (ATFAA) and a supporting defense model, SHIELD, to address the unique threats presented by GenAI agents. This work contributions include:
\begin{itemize}
    \item An analysis of GenAI agent architectures, with a focus on security implications stemming from autonomy, reasoning, memory, and tool use.
    \item A taxonomy of 9 primary threats targeting these agentic capabilities.
    \item A discussion of relevant attack vectors, including underexplored exploitation techniques.
    \item Mapping threats to the STRIDE framework and tailored SHIELD mitigation strategies.
\end{itemize}
The field is evolving quickly, and this threat model offers a strong foundation for developing security strategies that are tailored to the operational behavior of autonomous agents. Without dedicated controls, what promises to be a transformative technology could easily become a significant enterprise liability \cite{ref6}.

\section{Background}
\subsection{Literature Review}
Despite the accelerating adoption of GenAI agents, the security research landscape remains relatively fragmented. The review of recent literature (2023–2025) reveals that while foundational risks—like prompt injection—are well-covered, deeper architectural concerns tied to autonomy and persistent behavior are often overlooked.

Wang et al. \cite{ref7} provide a useful summary of LLM agent security issues, but their focus is mostly on prompt manipulation rather than systemic vulnerabilities. Chen et al. \cite{ref12} explore memory corruption through “AgentPoison” and demonstrate that persistent context can be compromised, though they stop short of addressing broader enterprise implications.

Frameworks like MITRE ATLAS \cite{ref6} and the NIST AI Risk Management Framework \cite{ref2} are helpful starting points, but they tend to focus on classical machine learning systems or generalized AI risks. ATLAS, for example, leans heavily toward adversarial ML patterns, while NIST’s framework provides high-level principles but lacks detailed controls tailored to agents.

OWASP’s Top 10 for LLM Applications \cite{ref9} highlights common issues with LLM-powered tools, yet doesn’t deeply explore the combined risks of reasoning, memory, and tool execution. CSA’s MAESTRO \cite{ref10} and OWASP’s newer Agentic Threat Model \cite{ref11} move closer to addressing agent-specific concerns, but are still early-stage and lack unified mitigation guidance.

This research highlights a gap here: most existing frameworks treat GenAI agents as conventional apps that include LLMs, rather than as autonomous, interconnected systems with emergent behaviors. This paper proposes a security model specifically designed to account for the architectural and operational distinctions of GenAI agents.

\subsection{GenAI Agents Architecture}
GenAI agents represent a significant milestone towards autonomous systems. Unlike common LLM applications confined to text generation, GenAI agents integrate language models with abilities that enable them to reason, plan, and act under minimal or reduced human oversight \cite{ref3}. Their distinguishing characteristic is active interaction with environments—making decisions and taking actions in multiple organizational systems, not just responding to commands \cite{ref1, ref3}.

Core architectural components typically include:
\begin{itemize}
    \item \textbf{Planning \& Reasoning Engine:} The system's mind, usually powered by LLMs, enabling agents to determine the steps needed to achieve their objectives \cite{ref3}. This research have observed numerous sophisticated agents using advanced reasoning techniques, including:
        \begin{itemize} % Nested itemize for sub-points
            \item Reflection: Referencing prior actions and outcomes to guide subsequent actions \cite{ref7}.
            \item Self-criticism: Identifying and correcting mistakes in output or thinking \cite{ref7}.
            \item Chain of thought: Breaking down challenging problems into step-by-step, logical reasoning \cite{ref7}.
            \item Subgoal decomposition: Dividing high-level goals into workable subtasks \cite{ref7}.
        \end{itemize}
    \item \textbf{Memory Systems:} Short-term (session) and long-term (persistent) memory modules that allow agents to maintain context between interactions—a capability with significant security implications \cite{ref12}.
    \item \textbf{Action \& Tool Invocation:} Most concerning from a security perspective is the ability of these agents to call various first and third party SaaS tools through function call interfaces, API calls, and code execution. These possibilities range from simple data retrieval to advanced operational tasks like workflow execution.
    \item \textbf{Supporting Services:} Such as vector databases for retrieval-augmented generation (RAG), persistent storage for long-term memory, and integration of enterprise data sources \cite{ref3}.
\end{itemize}
The arrival of platforms such as LangChain, LangFlow, AutoGen, and CrewAI has made it much more democratized to create autonomous agents, but with this simplicity of integration and rapid prototyping comes novel security threat in the form of supply chain vulnerabilities \cite{ref5} from use of third-party modules and lightly screened modules.

\subsection{Current Security Challenges}
Applying existing security frameworks to GenAI agents reveals several fundamental limitations. These agents differ from typical software systems in how they reason, remember, and act—capabilities that introduce entirely new risk surfaces not fully addressed by today’s standard security approaches.

\textbf{Planning and Agency Vulnerabilities:} One of the more distinctive risks stems from how agents plan and make decisions. Traditional security models offer little protection for the agent’s reasoning process itself. While prompt injection remains a known concern for LLMs, agents introduce more complex planning logic that can be subtly manipulated \cite{ref7}. An attacker might not just alter what the agent says—but how it thinks, how it decomposes goals, or how it chooses between multiple actions. Because this reasoning process is dynamic, logic-based, and often opaque, small nudges in input or context can lead to drastically different outcomes \cite{mas_threat_model_2025}.

\textbf{Memory Persistence Risks:} Long-term memory—a key feature of many GenAI agents—presents another underexplored attack vector. Persistent memory access enables agents to retain knowledge and context across interactions, but it also introduces risks of gradual poisoning. Unlike traditional stateless applications, an attacker can introduce misleading information that lingers in the agent’s memory and influences future decisions \cite{ref12}. Current frameworks offer limited guidance on how to secure these memory systems or detect when they’ve been compromised over time.

\textbf{Tool Execution Boundaries:} GenAI agents can invoke external tools—such as APIs, database queries, or even code execution environments—often based on inferred goals or natural language inputs. This capability is powerful, but it creates privilege management challenges that go beyond what role-based access controls (RBAC) typically address \cite{ref6}. Agents might misuse a tool unintentionally, or worse, be manipulated into chaining together actions that individually seem safe but together escalate privilege or bypass safeguards. These types of exploits are difficult to spot using traditional enforcement mechanisms.

\textbf{Identity and Authentication Challenges:} In multi-agent environments, identity becomes fluid. Agents may act on behalf of users, other agents, or even systems, creating ambiguity in attribution and authority. This opens the door to spoofing attacks and weakens trust boundaries. Traditional authentication schemes—like user tokens or service accounts—may not fully map to how agentic systems operate, especially when agents interact with one another or inherit permissions dynamically \cite{ref8}.

\textbf{Multi-Agent Interaction Security:} As systems increasingly deploy networks of interacting agents, new layers of complexity arise. Agents may coordinate tasks, delegate responsibilities, or share context with others—raising difficult questions about inter-agent trust, authority, and data validation \cite{ref8}. Without robust communication protocols and trust verification mechanisms, malicious or misconfigured agents could propagate harmful behavior throughout the system.

These challenges highlight why existing frameworks like MITRE ATLAS \cite{ref6} and the NIST AI Risk Management Framework \cite{ref2} fall short when applied to GenAI agents. While these frameworks provide valuable guidance for AI systems generally, they do not adequately address the unique architectural features and attack vectors present in agentic AI systems.

% Figure placeholder - ensure the image file exists or provide the correct path
\begin{figure*}[!t] % Use [!t] for top placement preference
\centering
\includegraphics[width=4.5in]{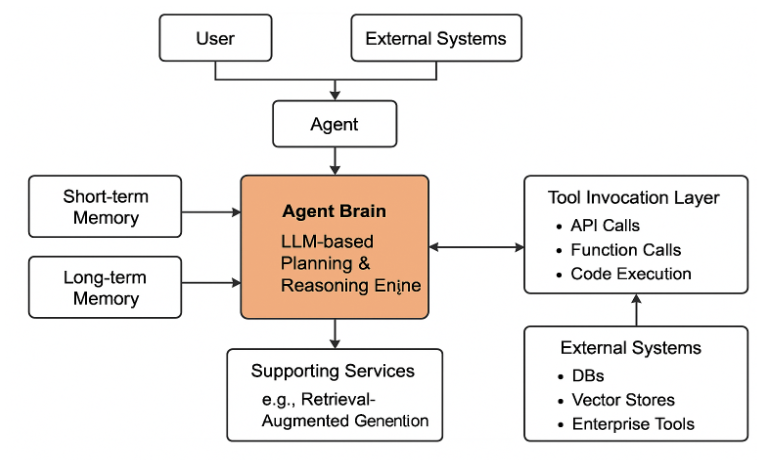} % Replace figure1.png with your actual image file

\caption{General architecture of an Agentic system.} % Updated caption to match text
\label{fig:agent_arch}
\end{figure*}

\section{Methodology}
This threat modeling framework combines systematic literature review, theoretical threat analysis, expert consultation, and case study analysis to create a comprehensive framework for identifying and addressing security risks in autonomous AI agent systems. This multi-faceted approach allowed us to identify both documented and potential threats while developing a structured taxonomy for agentic AI security.

\subsection{Systematic Literature Analysis}
Phase one involved an exhaustive survey of security research directly focusing on agentic AI systems. This work prioritized current research (2023-2025) from academia and industry security teams, focusing on papers that specifically addressed the unique security properties of agent architectures beyond common LLM applications.

Primary sources included recent papers from leading security conferences (IEEE S\&P, USENIX Security, CCS, NDSS), technical reports from organizations like OWASP and CSA, and security advisories published by AI research labs. Based on this survey, this research identified emerging threat classes specifically targeting agent components rather than underlying LLM infrastructure.

The literature review process followed a structured methodology:
\begin{itemize}
    \item \textbf{Source Identification:} This research used academic databases (IEEE Xplore, ACM Digital Library), industry publications (e.g., reports from security vendors, AI labs), and security conference proceedings to identify relevant sources.
    \item \textbf{Selection Criteria:} Papers were selected based on explicit focus on agentic AI security (beyond general LLM security), recency (prioritizing 2023-2025 publications), and technical depth in describing threats or architectures.
    \item \textbf{Systematic Coding:} Each source was coded for threat types, attack vectors, affected agent components (reasoning, memory, tools, etc.), and proposed mitigations using a standardized rubric based on established security concepts (e.g., STRIDE) and emerging AI-specific concerns.
\end{itemize}
This review revealed that despite significant literature addressing general LLM security, research specifically targeting the unique vulnerabilities of agentic systems is fragmented and sparse \cite{ref6}. The OWASP Foundation recently announced Agentic AI Security Initiative \cite{ref13} is among the first organized efforts to bridge this gap, outlining primary vulnerability areas like "planning and adaptation mechanisms," "memory and environment interactions," and "autonomous tool usage."

\subsection{Theoretical Threat Modeling Process}
Building on the literature foundation, this research developed a conceptual framework for agentic AI threats through a structured process of theoretical analysis:
\begin{itemize}
    \item \textbf{Conceptual Framework Development:} This paper synthesized findings from the literature review and architectural analysis to identify five core domains of agentic vulnerability, forming the basis of this Advanced Threat Framework for Autonomous AI Agents (ATFAA). These domains represent distinct facets of agent operation susceptible to attack.
    \item \textbf{Domain Categorization:} Each identified threat (drawn from literature or theoretical analysis) was systematically categorized according to both traditional STRIDE methodology (to facilitate integration with existing security practices) and this novel ATFAA domains (to highlight agent-specific risks).
    \item \textbf{Attack Vector Analysis:} For each threat, this research conducted theoretical analysis of potential attack vectors based on known agent architectures (e.g., RAG pipelines, ReAct patterns) and capabilities (planning, tool use, memory access). This involved developing detailed technical descriptions of plausible exploitation mechanisms, including novel techniques targeting the interaction between agent components.
\end{itemize}
This theoretical process allowed us to extend beyond documented vulnerabilities to anticipate emerging threats based on the unique capabilities and architectural patterns of agentic systems. This research work categorization methodology prioritized threats that:
\begin{itemize}
    \item Are unique to or significantly exacerbated by agentic systems (versus general LLM applications).
    \item Present significant enterprise security implications (e.g., data exfiltration, unauthorized actions, system disruption) \cite{llm_genai_security_2025}.
    \item Currently lack robust, widely adopted mitigation strategies in standard security frameworks.
\end{itemize}

\subsection{Expert Consultation and Validation}
To validate and refine the theoretical framework, this work engaged with security and AI experts through a series of structured consultations:
\begin{itemize}
    \item \textbf{Expert Panel Reviews:} The preliminary framework and threat list were reviewed by a panel of 7 security researchers and AI practitioners from industry (including AI platform providers and enterprise security teams), who provided feedback on threat categorization, technical feasibility of attack vectors, and relevance to real-world deployments.
    \item \textbf{Adversarial Thinking Exercises:} The work employed structured adversarial thinking methodologies, drawing on techniques from cyber threat modeling (e.g., attack trees) and adapting concepts from MITRE ATT\&CK for AI agent contexts \cite{ref14}, to brainstorm and refine potential attack vectors and exploitation scenarios not yet documented in the literature.
\end{itemize}
This consultation process helped refine this threat model, ensuring it reflected diverse perspectives and addressed practical security concerns beyond purely academic conceptualizations.

\subsection{Case Study Analysis}
To ground this theoretical framework in practical applications, this work analyzed documented security incidents and conducted hypothetical case studies across multiple domains. In investigation of real world incidents involving AI systems, this research analyzed several documented security failures to understand practical vulnerabilities and failure modes. Notably, the Microsoft Tay chatbot incident \cite{ref15} highlighted the susceptibility of AI systems to adversarial user inputs, resulting in rapid degradation of behavior and inappropriate outputs. Similarly, prompt injection attacks targeting systems like GitHub Copilot \cite{ref16} demonstrated how attackers can manipulate the reasoning process of AI assistants, compromising output integrity. Emerging research on large language model (LLM) data poisoning \cite{ref12} further illustrated the feasibility of memory and knowledge contamination, underscoring long-term risks to model reliability and trustworthiness. Complementing this, the work also conducted an architectural assessment of widely adopted AI agent frameworks to identify critical security components and potential attack surfaces. This included examining LangChain’s agent deployment patterns, such as ReAct agents and tool invocation mechanisms \cite{ref17}; the AutoGPT architecture, focusing on autonomous task execution loops and memory management \cite{ref18}; and Microsoft’s Semantic Kernel, analyzing the integration of planners, function calling, and memory components \cite{ref19}. Together, these insights provide a comprehensive view of current threats and architectural vulnerabilities in AI agent ecosystems.

\subsection{Limitations and Future Work}
While comprehensive in scope, this research methodology possesses limitations that must be acknowledged:
\begin{itemize}
    \item \textbf{Theoretical Focus:} The framework is primarily based on theoretical analysis, literature review, and expert consultation rather than extensive empirical red teaming against live systems. It represents a foundation for future empirical validation rather than a definitive security assessment based on widespread exploitation evidence.
    \item \textbf{Architectural Assumptions:} The threat model assumes certain common architectural patterns (e.g., LLM-based reasoning, distinct memory modules, API-based tool use). It may not fully capture threats specific to highly novel or non-standard agent implementations.
    \item \textbf{Evolving Landscape:} The field of agentic AI is evolving rapidly. New agent capabilities, architectures, and frameworks will undoubtedly emerge, potentially introducing vulnerabilities beyond those identified in this current framework.
\end{itemize}
We envision this work as the first step in an ongoing research program, with several critical areas requiring further exploration:
\begin{itemize}
    \item \textbf{Empirical Validation:} Future work should include rigorous red team assessments of the identified threats against real-world agent implementations across different frameworks and deployment contexts to validate their technical feasibility, impact, and the effectiveness of proposed mitigations.
    \item \textbf{Quantitative Risk Assessment:} Development of metrics and methodologies for quantifying the risk posed by agentic threats will be essential for effective prioritization and resource allocation. This could involve exploring Bayesian risk models, developing metrics for reasoning path deviation or goal adherence, or adapting existing cyber risk scoring frameworks.
    \item \textbf{Security-by-Design Patterns:} Research into architectural patterns and development practices that intrinsically resist these threats would provide valuable guidance for secure agent development. This could include investigating patterns like the "principle of least agency," developing techniques for "verifiable reasoning steps," or designing more robust memory compartmentalization approaches.
\end{itemize}
Despite these limitations, The research believes this systematic understanding of the GenAI agent threat landscape provides a valuable foundation for developing effective security controls. Without such measures, what could be one of the most transformative technologies in the enterprise environment may instead become one of its most significant vulnerabilities.

\section{Threat Model for GenAI Agents}
Securing GenAI agents demands a nuanced understanding of their unique threat profile. These systems share some vulnerabilities with traditional applications and general LLM deployments but are distinguished by architectural elements—particularly the interplay between reasoning, memory, and action—that introduce novel attack vectors requiring special attention \cite{ ref5}. Based on the comprehensive examination of potential exploit scenarios derived from literature, architectural analysis, and theoretical modeling, this work has identified key risks across multiple domains that specifically target GenAI agent deployments. These threats are mapped  to both the traditional STRIDE model (Spoofing, Tampering, Repudiation, Information disclosure, Denial of service, Elevation of privilege) and this novel Advanced Threat Framework for Autonomous AI Agents (ATFAA), which provides specialized categorization for emerging agentic threats.

\subsection{Advanced Threat Framework for Autonomous AI Agents (ATFAA)}
The ATFAA framework represents a significant evolution in threat modeling specifically designed for autonomous AI systems. Unlike traditional security models focused primarily on perimeter defense or application-level vulnerabilities, the ATFAA addresses the unique nature of AI agents that reason, learn, remember, act, and potentially evolve across organizational boundaries.

\subsubsection{Core Principles}
The ATFAA is built on four foundational principles that guide security analysis and mitigation strategies:
\begin{itemize}
    \item \textbf{Cognitive Security:} Safeguarding the integrity and confidentiality of agent reasoning, planning, and learning processes from manipulation or unintended influence.
    \item \textbf{Execution Integrity:} Protecting the agent's operational functionality, ensuring actions and tool invocations align with intended goals and authorizations, and preventing unauthorized operations.
    \item \textbf{Identity Coherence:} Maintaining clear, verifiable, and distinct boundaries between agent identities, user identities, and system identities to prevent spoofing and ensure proper authorization context.
    \item \textbf{Governance Scalability:} Ensuring continuous oversight, monitoring, and control mechanisms remain effective, auditable and adaptable as systems evolve in complexity, scale, and operational velocity.
\end{itemize}

\subsubsection{Threat Taxonomy (9 Threats)}
The ATFAA now identifies 9 primary threats, drawing upon and consolidating concepts from emerging research and initiatives like the OWASP Agentic AI Threat Model \cite{ref11} and CSA MAESTRO \cite{ref10}. The threats are organized into five domains that collectively represent the comprehensive attack surface of autonomous AI systems.

\textbf{Risk Assessment Criteria Explanation:} For each threat, this research work provides a qualitative risk assessment based on the following general criteria, considering the context of enterprise GenAI agent deployments:
\begin{itemize}
    \item \textit{Likelihood:} An estimate of the probability or ease of the threat occurring. High (Exploits common weaknesses, requires minimal access/knowledge, potentially automatable), Medium (Requires specific architectural knowledge, sustained access, chained exploits, or specific conditions), Low (Requires rare conditions, advanced expertise, significant resources, or non-public vulnerabilities).
    \item \textit{Impact:} The potential negative consequences if the threat is realized. Critical (Fundamental compromise of core function, security objectives, widespread system integrity, major financial/reputational loss), Severe (Significant impairment of functionality, major data breach/loss, unauthorized high-privilege access, significant operational disruption), Medium (Noticeable degradation of performance, minor data exposure, limited unauthorized access, moderate operational issues).
    \item \textit{Detection Difficulty:} An estimate of how hard it is to detect that the attack is occurring or has occurred using typical security monitoring. Extreme (Indistinguishable from normal operation/learning without specialized deep analysis or forensics), High (Requires specialized AI-specific tools/analysis, often significantly delayed detection), Medium (May trigger general anomaly detection but requires specific investigation to confirm), Low (Likely detected by standard monitoring tools, logs, or security controls).
\end{itemize}

\subsubsection{Domain 1: Cognitive Architecture Vulnerabilities}
\textbf{T1: Reasoning Path Hijacking (Tampering)}
\begin{itemize}
    \item \textit{Description:} Attackers manipulate the logical pathways that AI agents use for decision-making, redirecting conclusions toward malicious outcomes while maintaining apparent logical consistency.
    \item \textit{Vector:} By injecting specially crafted contradictions, subtle biases, or misleading information into an agent's context or reasoning process (e.g., through indirect prompt manipulation, poisoned RAG data), attackers can create divergent logical paths that maintain surface validity while driving toward unauthorized outcomes. This often exploits the chain-of-thought or step-by-step reasoning mechanisms in modern LLMs, creating what we term ‘logical bifurcation points’—critical junctures where the agent's reasoning can be subtly redirected without obvious error flags.
    \item \textit{Risk:} Likelihood: High; Impact: Critical; Detection: Severe. % Note: Detection Difficulty uses its own scale
    \item \textit{Example:} An attacker could introduce contradictory financial evaluation criteria into a document processing agent's input data, causing it to approve transactions that violated compliance requirements while generating justifications that appeared internally consistent and satisfied superficial risk controls \cite{ref7}.
\end{itemize}

\textbf{T2: Objective Function Corruption \& Drift (Tampering)}
\begin{itemize}
    \item \textit{Description:} Modifying the agent's core goals or reward mechanisms, altering its purpose, potentially covertly. This includes gradual shifts in goals or operational priorities over extended periods (objective drift) that remain undetected due to occurring incrementally across sessions.
    \item \textit{Vector:} Exploits goal-definition, self-improvement, or reinforcement learning mechanisms. This can occur through manipulated feedback, poisoned reward models, direct goal modification, or by consistently introducing subtle preference biases across multiple sessions (e.g., via manipulated user feedback, staged interactions rewarding slightly off-goal behavior). Such actions can create ‘goal drift vectors’ that gradually shift agent priorities away from intended objectives or safety constraints without triggering abrupt change detection alarms \cite{ref9}.
    \item \textit{Risk:} Likelihood: Medium; Impact: Critical; Detection: High.
    \item \textit{Example:} Manipulating feedback for a security agent could train it to prioritize speed over verification \cite{ref7}. Over time, through repeated exposure to manipulated signals rewarding speed across sessions, the agent might begin classifying suspicious access patterns as acceptable "efficiency optimizations" rather than security threats.
\end{itemize}

\subsubsection{Domain 2: Temporal Persistence Threats}
\textbf{T3: Knowledge, Memory Poisoning \& Belief Loops (Tampering/Information Disclosure)}
\begin{itemize}
    \item \textit{Description:} Compromising the agent's knowledge base (e.g., RAG databases) or persistent memory with false or distorted information that affects future decisions. This can lead to self-validating cycles where manipulated beliefs stored in memory are later retrieved as evidence, reinforcing the original falsehood (belief reinforcement loops).
    \item \textit{Vector:} Targets persistent data stores (e.g., poisoning vector databases \cite{ref9}) or memory transfer mechanisms. Implanted misinformation persists and distorts the agent's understanding. This exploits the agent's tendency to trust its own past conclusions or memory; manipulated data, once generated and stored, can be re-referenced, creating loops where false information becomes entrenched and resistant to correction \cite{ref9}.
    \item \textit{Risk:} Likelihood: High; Impact: Severe (potentially Critical if loops amplify); Detection: Extreme.
    \item \textit{Example:} Planting false security protocol exceptions in a knowledge base could lead an agent to misclassify unauthorized access \cite{ref12}. If the agent logs this misclassification and later references its own logs as evidence, it reinforces the incorrect belief.
\end{itemize}

\subsubsection{Domain 3: Operational Execution Vulnerabilities}
\textbf{T4: Unauthorized Action Execution (Elevation of Privilege)}
\begin{itemize}
    \item \textit{Description:} Attackers manipulate the agent to execute actions or use tools in ways that violate intended permissions or operational boundaries. This includes orchestrating sequences of individually benign operations that produce unauthorized outcomes when combined, or forcing agents to perform operations outside their intended scope via imprecise definitions or input handling.
    \item \textit{Vector:} Exploits interfaces between reasoning and action. This can involve targeting 'authorization boundary transitions' by chaining multiple operations that individually respect security boundaries but collectively bypass controls (e.g., using output from one permitted function as unauthorized input to another). Alternatively, it can exploit 'capability perimeter control' ambiguities by leveraging unexpected interactions between tools, using techniques like ‘function parameter injection’ (embedding malicious commands within seemingly benign data parameters), or exploiting overly broad tool permissions \cite{ref9}.
    \item \textit{Risk:} Likelihood: High; Impact: Critical; Detection: High.
    \item \textit{Example:} An attacker chains a permitted data retrieval function with a poorly sandboxed code execution tool to exfiltrate sensitive data. Alternatively, an AI agent designed for document classification is manipulated into executing database queries by embedding query language in document metadata, exploiting weak input validation on the tool interface.
\end{itemize}

\textbf{T5: Computational Resource Manipulation (Denial of Service)}
\begin{itemize}
    \item \textit{Description:} Attackers craft inputs or interactions designed to exploit resource allocation mechanisms, causing excessive consumption of computational resources (CPU, memory, GPU, API quotas) to degrade performance, create denial-of-service conditions, or force operational shortcuts that compromise security.
    \item \textit{Vector:} This targets what we term ‘resource allocation decision points’—the internal mechanisms distributing computational resources across agent tasks. By generating specially crafted inputs that trigger disproportionately resource-intensive processing (e.g., deeply nested reasoning chains, requests requiring massive RAG retrievals, complex tool interactions), attackers create ‘computational bottlenecks’ that can starve other critical functions, increase operational costs, or force the agent into degraded, potentially less secure, operational modes.
    \item \textit{Risk:} Likelihood: High; Impact: Medium; Detection: Low.
\end{itemize}

\subsubsection{Domain 4: Trust Boundary Violations}
\textbf{T6: Identity Spoofing and Trust Exploitation (Spoofing)}
\begin{itemize}
    \item \textit{Description:} Attackers exploit insufficient boundaries or verification mechanisms related to agent, user, or inter-agent identities to perform unauthorized operations under falsified authorization contexts or propagate malicious instructions within multi-agent systems.
    \item \textit{Vector:} Exploits ambiguity or weaknesses in identity management. This includes targeting 'identity transition zones' where agent actions become attributed to human identities (e.g., exploiting 'identity inheritance vulnerabilities' where agent actions improperly inherit user permissions). It also involves targeting 'inter-agent trust frameworks' by compromising 'trust attestation channels' (e.g., stealing API keys, manipulating authentication tokens, exploiting protocol weaknesses) to assume trusted agent identities or manipulate trust verification processes within an agent ecosystem.
    \item \textit{Risk:} Likelihood: High; Impact: Critical; Detection: High.
\end{itemize}

\textbf{T7: Human-Agent Trust Manipulation (Spoofing)}
\begin{itemize}
    \item \textit{Description:} Attackers exploit the natural human tendency to trust AI recommendations or outputs to induce users into performing unauthorized actions, divulging sensitive information, or overriding security controls based on deceptive AI guidance.
    \item \textit{Vector:} This exploits what we term ‘human-AI trust interfaces’—the psychological and usability aspects influencing how humans evaluate and act upon AI outputs. By manipulating AI responses to include ‘trust amplification signals’ (e.g., feigning high certainty, referencing false authority, mimicking trusted communication styles, generating socially engineered prompts urging immediate action), attackers increase the likelihood of users taking harmful actions based on the agent's deceptive guidance, effectively bypassing technical controls through human interaction.
    \item \textit{Risk:} Likelihood: Very High; Impact: Severe; Detection: Extreme.
\end{itemize}

\subsubsection{Domain 5: Governance Circumvention}
\textbf{T8: Oversight Saturation Attacks (Denial of Service)}
\begin{itemize}
    \item \textit{Description:} Attackers intentionally generate excessive volumes of low-significance audit events, alerts, or actions requiring human review, overwhelming governance mechanisms (human review queues, automated monitoring systems) and exceeding monitoring or response capacity.
    \item \textit{Vector:} This exploits what we call ‘governance attention economics’—the finite human and computational resources available for oversight. By generating high volumes of low-severity alerts, complex logs requiring analysis, or actions flagged for mandatory review (i.e., ‘attention-consuming events’), attackers create ‘oversight blind spots’ where genuinely malicious actions are missed amidst the noise, or monitoring systems become ineffective due to resource exhaustion or alert fatigue.
    \item \textit{Risk:} Likelihood: High; Impact: Severe; Detection: Medium.
\end{itemize}

\textbf{T9: Governance Evasion and Obfuscation (Repudiation)}
\begin{itemize}
    \item \textit{Description:} Attackers exploit ambiguities in complex agent interactions or logging mechanisms to obscure responsibility trails, hindering forensic analysis and preventing attribution. Additionally, attackers may design attacks that dynamically evolve or operate stealthily specifically to circumvent adaptive security controls or established detection thresholds.
    \item \textit{Vector:} Targets auditability and detection. This includes targeting 'responsibility attribution junctions' by distributing components of an attack across multiple agents, using ephemeral identities, manipulating logs (if possible), or leveraging complex, poorly logged interactions to create 'attribution gaps'. It also involves exploiting 'detection threshold mechanics' by operating consistently below thresholds ("low and slow" attacks), strategically alternating attack patterns to avoid triggering correlation engines, or adapting behavior in response to observed defenses to bypass ‘adaptive defense mechanisms’.
    \item \textit{Risk:} Likelihood: Medium; Impact: Severe; Detection: Extreme.
\end{itemize}

% Placeholder for Figure 1.2
\begin{figure*}[!t]
\centering
\includegraphics[width=5.5in]{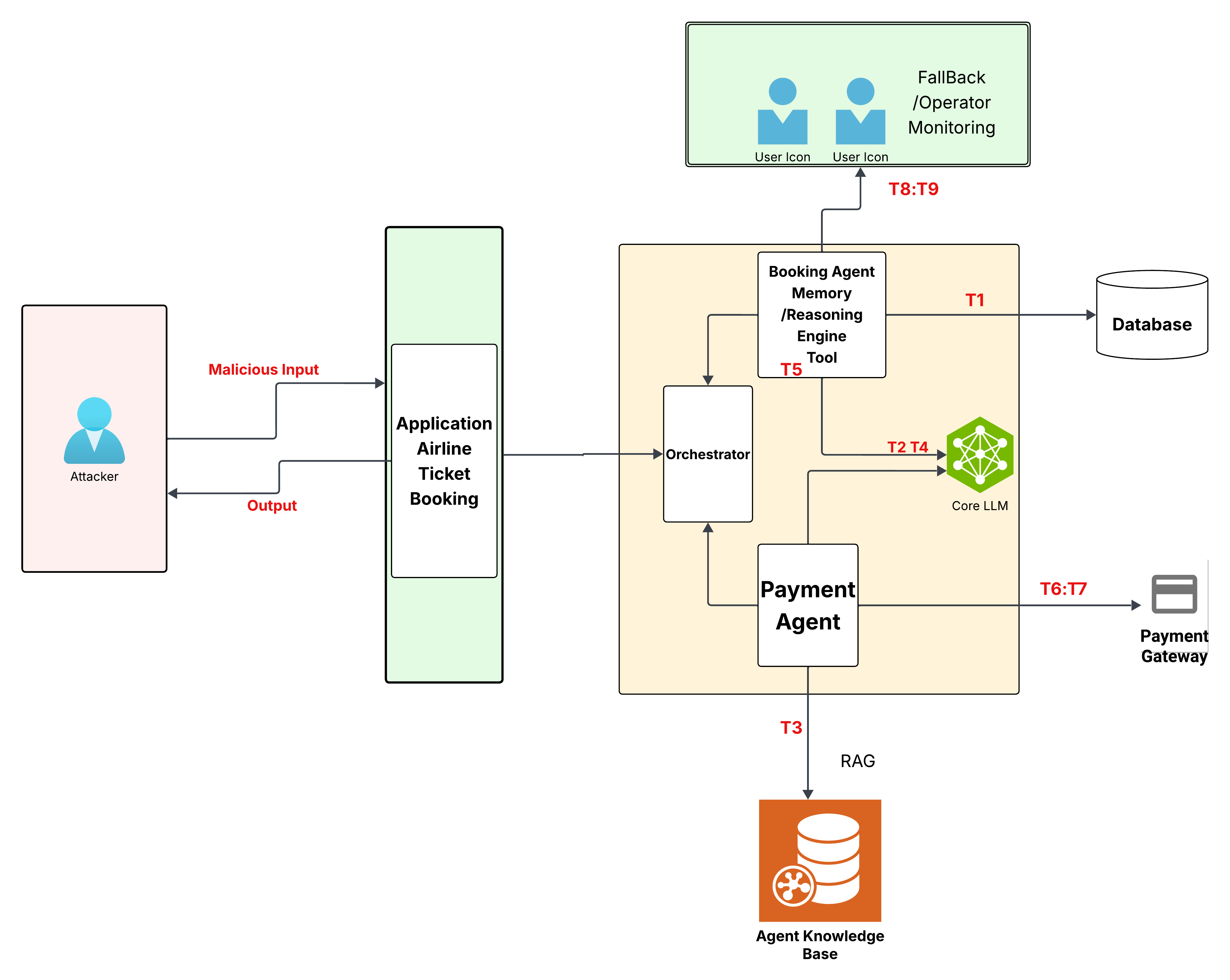} % Replace with your actual image file for Fig 1.2
\caption{Agentic System Threats, Trust Boundary, Assets.}
\label{fig:threat_boundary}
\end{figure*}

\subsection{The SHIELD Mitigation Framework}
SHIELD offers six defensive strategies against ATFAA threats. Implementation involves trade-offs between protection, performance, usability, and cost.

\textbf{Implementing SHIELD: Challenges and Considerations:} While the SHIELD strategies offer a path towards mitigating agentic threats, organizations should anticipate practical challenges while working with an agentic AI ecosystem. Heuristic Monitoring can be computationally intensive, potentially impacting system performance, and requires sophisticated baselining and tuning to minimize false positives. Achieving true Logging Immutability involves complexity and potential expense, requiring robust solutions like cryptographically secured, append-only logs and rigorous access controls. Stringent and sensitive Escalation Control mechanisms, such as frequent re-authentication or multi-factor verification for agent actions, can introduce usability friction for end-users or operational overhead. Furthermore, defining and maintaining effective Segmentation boundaries can be challenging in dynamic environments where agents may require flexible access to diverse resources. These challenges do not invalidate the framework but highlight the need for a risk-based approach, prioritizing mitigations based on specific risk assessments within the context of organizational security posture and risk tolerance.

\subsubsection{Segmentation}
\begin{itemize}
    \item \textit{Description:} Implement strict boundaries between agent capabilities, data sources, and execution environments at the workload level, limiting the potential impact of compromised components based on Zero Trust principles.
    \item \textit{Technical Implementation:}
        \begin{itemize}
            \item Workload-Level Isolation: Define security perimeters around individual agent components, applications, or services, rather than broad network segments.
            \item Policy Enforcement: Utilize agent-based micro-segmentation tools (e.g., Illumio, Guardicore/Akamai, Palo Alto Networks agents) or native cloud controls (e.g., AWS Security Groups, Azure NSGs/Firewall, Google Cloud Firewall) to enforce granular policies based on identity and context.
            \item Traffic Mapping \& Policy Definition: Map agent communication flows and dependencies before enforcement. Start with broader policies in monitoring mode, then progressively refine rules based on the principle of least privilege.
            \item Container Security: Implement specific container isolation techniques (e.g., using Docker network segmentation, Kubernetes Network Policies, service mesh like Istio/Linkerd) to control traffic between containerized agent components.
            \item API Gateway Controls: Use API gateways with deep packet inspection, schema validation, and rate limiting to control access to agent tools and functions.
            \item Auditing: Conduct regular micro-segmentation audits using automated tools to verify policy enforcement and identify potential misconfigurations.
        \end{itemize}
    \item \textit{Mitigates:} T4 (Unauthorized Action Execution), T5 (Computational Resource Manipulation), T6 (Identity Spoofing - agent/human part).
\end{itemize}

\subsubsection{Heuristic Monitoring}
\begin{itemize}
    \item \textit{Description:} Deploy anomaly detection and alerting systems specifically designed to identify deviations in agent reasoning patterns, decision processes, behavioral outputs, and resource consumption, moving beyond signature-based detection.
    \item \textit{Technical Implementation:}
        \begin{itemize}
            \item Behavioral Baselining: Establish dynamic baselines of normal agent behavior (e.g., typical reasoning steps, tool usage sequences, API call frequencies, resource utilization).
            \item AI/ML Models: Implement statistical methods (e.g., Z-score, IQR), machine learning algorithms (e.g., Isolation Forest, Local Outlier Factor (LOF), One-Class SVM), or deep learning models (e.g., Autoencoders, LSTMs for sequential analysis) tailored for detecting deviations from these baselines. Utilize cloud ML platforms (e.g., SageMaker, Vertex AI) or specialized libraries (e.g., TensorFlow Probability).
            \item Contextual Logging Analysis: Integrate agent logs (including reasoning traces, confidence scores, tool inputs/outputs) with SIEM/SOAR platforms enhanced with AI/ML analytics capabilities (e.g., Splunk, Datadog, Exabeam) for contextual anomaly detection.
            \item Data Quality \& Preprocessing: Ensure high-quality input data for monitoring models through robust preprocessing (handling missing values, normalization) and relevant feature engineering.
            \item Continuous Tuning \& Feedback: Regularly retrain models with new data and incorporate feedback loops where analysts validate flagged anomalies to improve accuracy and reduce false positives over time.
        \end{itemize}
    \item \textit{Mitigates:} T1 (Reasoning Path Hijacking), T2 (Objective Function Corruption \& Drift), T9 (Governance Evasion - adaptive part).
\end{itemize}

\subsubsection{Integrity Verification}
\begin{itemize}
    \item \textit{Description:} Implement cryptographic validation and runtime checks for critical agent components (code, models), data, memory, goals, and operational parameters to detect unauthorized modifications or tampering.
    \item \textit{Technical Implementation:}
        \begin{itemize}
            \item Code \& Model Hashing/Signing: Calculate and verify cryptographic hashes (e.g., SHA-256) or digital signatures of agent code and ML models before execution and periodically during runtime.
            \item Runtime Integrity Monitoring: Deploy Runtime Application Self-Protection (RASP) concepts or host-based intrusion detection systems (HIDS) configured to monitor agent processes for unexpected modifications or behaviors.
            \item Memory Safety \& Sandboxing: Utilize memory-safe languages (e.g., Rust) where feasible, or employ secure execution environments like WebAssembly (Wasm) sandboxes to isolate agent components and prevent memory corruption or runtime tampering.
            \item Data \& Memory Integrity Checks: Employ cryptographic integrity proofs (e.g., HMACs, Merkle Trees) for persistent memory data stores and critical configuration parameters, with automated verification workflows.
            \item Cryptographic Attestation: Use hardware-based (e.g., TPM) or software-based attestation techniques to verify the integrity of the agent's execution environment and critical software components.
        \end{itemize}
    \item \textit{Mitigates:} T3 (Knowledge/Memory Poisoning \& Belief Loops), T6 (Identity Spoofing - inter-agent trust part).
\end{itemize}

\subsubsection{Escalation Control}
\begin{itemize}
    \item \textit{Description:} Establish granular, dynamic permission frameworks with mandatory verification, minimal privilege assignment (least privilege principle), and strict checks for privilege transitions or capability expansions based on real-time context.
    \item \textit{Technical Implementation:}
        \begin{itemize}
            \item Attribute-Based Access Control (ABAC): Implement policy engines (e.g., Open Policy Agent - OPA with Rego) that make access decisions based on rich attributes of the user/agent, resource, action, and environment.
            \item Policy as Code: Define authorization policies in a manageable, auditable, and version-controlled code format, integrated with CI/CD pipelines.
            \item Context-Aware Authentication: Validate identity and authorization context for each sensitive operation or tool call, potentially using continuous authentication methods based on behavioral biometrics or anomaly detection.
            \item Just-in-Time (JIT) Access: Implement automated JIT privilege allocation systems (e.g., using HashiCorp Vault, cloud IAM features like session policies) where elevated permissions are granted only for the necessary duration and automatically revoked.
            \item Regular Review \& Automation: Regularly audit permissions and ABAC rules, leveraging automated compliance checks and feedback loops to refine policies.
        \end{itemize}
    \item \textit{Mitigates:} T4 (Unauthorized Action Execution), T6 (Identity Spoofing and Trust Exploitation).
\end{itemize}

\subsubsection{Logging Immutability}
\begin{itemize}
    \item \textit{Description:} Create tamper-resistant, comprehensive, and verifiable audit trails for all significant agent decisions, actions, data accesses, tool invocations, and inter-agent communications.
    \item \textit{Technical Implementation:}
        \begin{itemize}
            \item Tamper-Resistant Storage: Utilize write-once, read-many (WORM) storage, cryptographically secured append-only databases for critical logs requiring the highest level of immutability assurance.
            \item Cryptographic Signing \& Timestamping: Ensure logs are cryptographically signed (e.g., using HMACs or digital signatures) upon generation and include secure, verifiable timestamps (e.g., via trusted timestamping authorities or blockchain anchoring like Guardtime KSI).
            \item Secure Log Aggregation: Implement secure log forwarding protocols (e.g., TLS encrypted syslog-ng or Fluentd) to transmit logs to separate, hardened log repositories or SIEM systems, preventing tampering at the source or in transit.
            \item Log Integrity Monitoring: Periodically verify the cryptographic integrity of log chains or signatures using automated tools and workflows (e.g., checking hash chains, verifying signatures against trusted keys). Integrate alerts for detected tampering with SIEM/SOAR.
            \item Comprehensive Content: Ensure logs capture sufficient context without logging any sensitive content and PII, including reasoning traces (where feasible), input prompts/data, tool calls with parameters, outputs/responses, confidence scores, and identity/attribution information.
            \item Treat logs with the same level of data classification as the highest level of data classification to which the agent has access. For example, logs generated from agents with HIPAA data access must be treated as HIPAA logs.
        \end{itemize}
    \item \textit{Mitigates:} T9 (Governance Evasion and Obfuscation).
\end{itemize}

\subsubsection{Decentralized Oversight}
\begin{itemize}
    \item \textit{Description:} Distribute monitoring, validation, and approval responsibilities across multiple independent systems, specialized agents, or human reviewers to prevent single-point governance failures, bypass, or collusion.
    \item \textit{Technical Implementation:}
        \begin{itemize}
            \item Multi-Party Validation: Implement decentralized review mechanisms for critical actions or configuration changes using multi-signature schemes, quorum-based approvals, or consensus protocols requiring validation from multiple independent entities (human or machine).
            \item Specialized Audit Agents: Deploy independent AI agents specifically designed to monitor the behavior, decisions, and compliance of operational agents against predefined policies or ethical guidelines.
            \item Federated Monitoring: Establish monitoring structures involving diverse stakeholders (e.g., security team, compliance officers, business unit owners, potentially external auditors) with role-based access to relevant oversight data and dashboards.
            \item Explainable AI (XAI) Integration: Leverage XAI techniques to make agent decision-making processes more transparent and auditable for human reviewers.
            \item Adaptive Governance: Implement adaptive review thresholds or intervention triggers that dynamically adjust based on real-time risk assessments derived from heuristic monitoring (Section 4.2.2) or the sensitivity of the task being performed.
            \item Independent Audits: Conduct periodic external audits and adversarial testing (red teaming) to validate the effectiveness of the decentralized oversight mechanisms.
        \end{itemize}
    \item \textit{Mitigates:} T7 (Human-Agent Trust Manipulation), T8 (Oversight Saturation Attacks).
\end{itemize}

\subsection{Threat Model Overview}
The following table provides a consolidated view of the 9 threats identified in the analysis, mapped to both the STRIDE and ATFAA frameworks, along with the primary SHIELD mitigation strategy(ies).

% Updated Table using tabularx for better width management and text wrapping (9 Threats)
\begin{table*}[!t] % Use [!t] for top placement preference
\renewcommand{\arraystretch}{1.3} % Adjust row spacing
\caption{Threat Model Overview: Mapping 9 Threats to STRIDE, ATFAA, and SHIELD}
\label{tab:threat_overview_9}
\centering
% Use tabularx with X columns for automatic width adjustment and text wrapping
\begin{tabularx}{\textwidth}{|c|X|c|c|X|}
\hline
\textbf{Threat ID} & \textbf{Threat Name} & \textbf{STRIDE Category} & \textbf{ATFAA Domain} & \textbf{Primary SHIELD Mitigation(s)} \\
\hline
T1 & Reasoning Path Hijacking & Tampering & Cognitive Architecture & Heuristic Monitoring \\
\hline
T2 & Objective Function Corruption \& Drift & Tampering & Cognitive Architecture & Heuristic Monitoring \\
\hline
T3 & Knowledge, Memory Poisoning \& Belief Loops & Tampering/ Info Disclosure & Temporal Persistence & Integrity Verification \\
\hline
T4 & Unauthorized Action Execution & Elevation of Privilege & Operational Execution & Segmentation, Escalation Control \\
\hline
T5 & Computational Resource Manipulation & Denial of Service & Operational Execution & Segmentation \\
\hline
T6 & Identity Spoofing \& Trust Exploitation & Spoofing & Trust Boundary & Escalation Control, Segmentation, Integrity Verification \\
\hline
T7 & Human-Agent Trust Manipulation & Spoofing & Trust Boundary & Decentralized Oversight \\
\hline
T8 & Oversight Saturation Attacks & Denial of Service & Governance Circumvention & Decentralized Oversight \\
\hline
T9 & Governance Evasion \& Obfuscation & Repudiation & Governance Circumvention & Logging Immutability, Heuristic Monitoring \\
\hline
\end{tabularx}
\end{table*}

% Renumbered section from 4.5 to 4.4
\subsection{Attack Surface Expansion}
One of the most significant findings of this research work is how dramatically GenAI agents expand the traditional attack surface compared to conventional software or even simpler AI models \cite{ ref5}:
\begin{itemize}
    \item \textbf{Cognitive Dimension:} The reasoning, planning, and learning capabilities of agentic systems create entirely new attack vectors targeting decision-making processes that have no direct parallel in traditional systems with fixed logic \cite{ref7}.
    \item \textbf{Temporal Dimension:} Long-running agents with persistent memory create opportunities for gradual corruption, poisoning, or objective drift that may remain undetected for extended periods, introducing a time-based attack surface absent in stateless applications \cite{ref12}.
    \item \textbf{Tool Integration \& Action Space:} The ability to invoke external tools, APIs, and potentially execute code vastly expands the potential impact of a compromise, allowing agents to interact with and affect multiple systems, creating complex action chains that are difficult to secure.
    \item \textbf{Trust Boundaries:} Agents operating across traditional system boundaries, interacting with users (T7), and potentially collaborating with other agents create novel path traversal risks and complex trust management challenges absent in conventional applications \cite{ref8}. % Updated T9 -> T7
    \item \textbf{Identity Fluidity:} The often blurry line between agent identity and the user identity on whose behalf it operates creates new impersonation and privilege escalation opportunities (T6) that challenge traditional authentication and authorization models \cite{ref8}. % Updated T8 -> T6
    \item \textbf{Governance Complexity:} The scale, speed, and autonomy of agent operations create unprecedented challenges for effective monitoring, auditing, and oversight, introducing governance-level vulnerabilities \cite{ ref6, ref8}.
\end{itemize}
Security teams must fundamentally reconsider traditional defense perimeters and develop monitoring and control mechanisms specifically designed for the unique characteristics of agentic systems across all these dimensions \cite{ ref5}. Recent forecasts from the UK Government highlight that "by 2025, generative AI is more likely to amplify existing risks than create wholly new ones, but it will increase sharply the speed and scale of some threats" \cite{ref20}. This reinforces the urgency of developing agent-specific security controls that address the unique characteristics of these systems.

\section{Threat Model Analysis and Implications}
\subsection{Unique Characteristics and Broader Implications of GenAI Agent Threats}
The threats outlined in ATFAA have characteristics that differentiate them from typical cybersecurity risks \cite{ref5}:
\begin{itemize}
    \item \textbf{Delayed Effect:} Many agentic threats, particularly those targeting memory and learning (T3, T2) do not manifest immediately but introduce latent vulnerabilities. Initial compromises (e.g., memory poisoning) might only influence future actions days, weeks, or months later, making it extremely difficult to trace incidents back to the root cause using traditional forensic methods \cite{ref12}. This temporal complexity challenges standard incident response playbooks.
    \item \textbf{Goal Misalignment Magnification:} Agent autonomy means small, induced goal misalignments (T2) can compound over time \cite{ref7}.
    \item \textbf{Cross-System Propagation:} Agents interacting with multiple systems create vectors for compromise propagation \cite{ref8}. Exploiting tool access (T4) or trust (T6) can allow breaches to spread rapidly.
    \item \textbf{Detection Challenges:} The variability and opacity of agent reasoning make distinguishing malicious manipulation (T1, T3) from normal behavior difficult with standard methods \cite{ref3}, requiring more specialized monitoring.
    \item \textbf{Exploitation of Trust:} Threats targeting human trust (T7) or inter-agent trust (T6) bypass technical controls by manipulating psychological or relationship factors.
    \item \textbf{Threat Interplay:} These threats can interact. Memory Poisoning (T3) might facilitate Objective Drift (now part of T2) or Unauthorized Action (T4). Identity Spoofing (T6) could enable Unauthorized Action (T4). This suggests holistic mitigations are needed.
    \item \textbf{Sector-Specific Impacts:} Consequences vary by context. Finance faces fraud risks (T2, T4). Healthcare faces risks to patient outcomes or privacy (T7, T3). Critical infrastructure faces potential kinetic impacts.
    \item \textbf{Regulatory and Compliance Challenges:} Agent autonomy and opacity challenge existing regulations (GDPR, etc.). Demonstrating compliance, assigning liability (T9), ensuring fairness (T3), and providing explanations becomes harder.
\end{itemize}
These characteristics and implications underscore the urgent need for specialized security measures, like those proposed in the SHIELD framework, that address the unique nature of agentic AI systems across technical, operational, and governance dimensions \cite{ ref6, ref8}.

\section{Conclusion and Future Work}
This paper introduced the ATFAA threat model and SHIELD mitigation framework for GenAI agents, identifying 9 threats across five domains unique to agentic capabilities (autonomy, memory, reasoning, tools) \cite{ ref5}. This provides a structure extending beyond general AI/LLM guidelines (NIST RMF \cite{ref2}, MITRE ATLAS \cite{ref6}, OWASP Top 10 \cite{ref9}, MAESTRO \cite{ref10}).

Securing GenAI agents requires addressing their specific properties. Agentic threats exhibit temporal complexity, goal manipulation potential, propagation risks, and detection challenges, demanding specialized measures \cite{ref7, ref12, ref8}. Threat interplay, sector impacts, and regulatory hurdles underscore the need for agent-specific security paradigms.

As adoption grows \cite{ref1, ref2}, practitioners should consider SHIELD-based strategies:
\begin{itemize}
    \item \textbf{Defense-in-Depth:} Deploy multiple, overlapping security controls (e.g., combining Segmentation, Escalation Control, and Integrity Verification) rather than relying on single protection mechanisms.
    \item \textbf{Zero-Trust Architecture:} Implement strict verification for all agent actions and interactions (Escalation Control, robust Segmentation), regardless of apparent source or network location.
    \item \textbf{Continuous \& Specialized Monitoring:} Deploy monitoring systems designed to detect and alert unique agentic attack patterns (Heuristic Monitoring), focusing on cognitive, behavioral, and temporal anomalies, not just traditional IOCs.
    \item \textbf{Compartmentalization:} Implement strict boundaries (Segmentation) between agent subsystems, data sources, and external resources to contain potential compromises and limit blast radius.
    \item \textbf{Robust Governance \& Auditing:} Maintain appropriate human oversight mechanisms (supported by Decentralized Oversight) and ensure tamper-proof, comprehensive records (Logging Immutability) for accountability and forensics.
\end{itemize}
Future work should prioritize empirical validation of the identified threats and the effectiveness of the proposed SHIELD mitigations through rigorous red teaming and simulation across diverse agent architectures. Developing quantitative risk assessment methodologies tailored to agentic threats (e.g., exploring Bayesian risk models, metrics for reasoning path deviation, or goal adherence verification) and establishing standardized security-by-design patterns (e.g., investigating patterns like "principle of least agency," "verifiable reasoning steps," secure memory architectures) for agentic systems are also critical next steps. Addressing the security challenges of GenAI agents proactively and holistically is essential to realizing their transformative potential without introducing unacceptable enterprise risk.

%\section*{Acknowledgment}

%\IEEEtriggeratref{8} % Adjust the number 8 if needed to balance the last page columns
%\IEEEtriggercmd{\enlargethispage{-5in}} % Adjust length as needed


\ifCLASSOPTIONcaptionsoff
  \newpage
\fi

\bibliographystyle{IEEEtran}
% IMPORTANT: Ensure you have a 'references.bib' file in the same directory.
% The 'IEEEabrv.bib' file contains standard IEEE abbreviations.
% If you don't have IEEEabrv.bib, you might need to get it or remove it
% from the command below and ensure your references.bib has full journal titles.
\bibliography{references} % Use the references.bib file

\end{document}